\documentclass[epj]{svjour}

\usepackage[T1]{fontenc}    
\usepackage{hyperref}       
\usepackage{url}            
\usepackage{booktabs}       
\usepackage{amsfonts}       
\usepackage{amsmath}
\usepackage{amssymb}
\usepackage{graphicx} 
\usepackage{graphics}
\usepackage{subcaption}
\usepackage{xcolor}
\begin{document}
%
\title{Observational constraints on Starobinsky $f(R)$ cosmology from cosmic expansion and structure growth data
}

\titlerunning{Observational constraints on Starobinsky $f(R)$ cosmology}

\author{P. Bessa \inst{1},  
M. Campista\inst{2,3}, 
A. Bernui\inst{3}
}                     
%
%
\institute{International PhD Program in
Astrophysics, Cosmology and Gravitation (PPGCosmo), Espírito Santo, Brazil \and 
Instituto Politécnico, Universidade Federal do Rio de Janeiro (UFRJ-Macaé), Macaé, Rio de Janeiro, Brazil \and 
Coordenação de Astronomia e Astrofísica, Observatório Nacional (ON), Rio de Janeiro, Brazil}
\date{Received: date / Revised version: date}
%
\abstract{
The unknown physical nature of the Dark Energy motivates in cosmology the study of modifications of the gravity theory at large distances. 
One of these types of  modifications is to consider gravity theories, generally termed as $f(R)$. 
In this paper we use observational data to both constrain and test the Starobinsky $f(R)$ model~\cite{Starobinsky2007}, using updated
measurements from the dynamics of the expansion of the universe, $H(z)$; and the growth rate of cosmic structures, $[f\sigma_8](z)$, 
where the distinction between the concordance $\Lambda $CDM model and modified gravity models $f(R)$ becomes clearer. 
We use MCMC likelihood analyses to explore the parameters space of the $f(R)$ model using $H(z)$ and $[f\sigma_8](z)$ data, both 
individually and jointly, and further, examine which of the models best fits the joint data. 
To further test the Starobinsky model, we use a method proposed by Linder~\cite{Linder2017}, where the data from the observables 
is jointly binned in redshift space. 
This allows to further explore the model's parameter that better fits the data in comparison to the $\Lambda$CDM model. 
The joint analysis of $H(z)$ and $[f\sigma_8](z)$ show that the $n=2$--Starobinsky $f(R)$ model fits well the observational data. 
In the end, we confirm that this joint analysis is able to break the degenerescence between modified gravity models 
as proposed in the original work~\cite{Starobinsky2007}. 
Our results indicate that the $f(R)$ Starobinsky model provides a good fit to the currently available data 
for a set of values of its parameters, being, therefore, a possible alternative to the $\Lambda$CDM model.
\PACS{ 
      {04.50.Kd} {Modified theories of gravity} \and
      {98.80.Es}{Observational cosmology}   \and
      {98.62.Py} {Distances, redshifts, radial velocities; spatial distribution of galaxies} \and
      {95.36.+x}{ Dark energy}
     } 
} 
\maketitle
%
\section{Introduction}

The search for an explanation to the  current phase of accelerated expansion of the universe is one of the most important paradigms in modern cosmology~\cite{Planck:2015bue}. 
With the available observational information, the concordance model of cosmology that best fits the data is the flat $\Lambda$CDM~\cite{Planck2018} model, 
where the universe is filled in with cold dark matter (CDM) and a dark energy (DE) component --in the form of a cosmological constant $\Lambda$-- 
in addition to the standard baryonic and electromagnetic ingredients. 
The main concern with the DE is that its physical interpretation is still unknown. 
Other plausible causes of the observed accelerated expansion should be
explored~\cite{Kazantzidis:2018rnb,Basilakos_2020,Linder_2020,Velasquez_Toribio_2020,Bonilla_2021,Davari_2021}. 
Because $\Lambda$CDM is based on Einstein's general relativity (GR), one possible way is to explore  cosmological models not based on GR~\cite{Linder,Benaoum,Perenon15,Perenon19,Perenon20,Batista14,Nunes18,Benisty18,Benisty19a,Benisty19b}.
However, there is a degeneracy problem between models supported by GR and models based on other metric theories, generally termed Modified Gravity 
(MG) theories. 
This degenerescence cannot be broken using data from the dynamical evolution of the Universe alone, as the same dynamics can be explained by different MG theories as well as by evolving DE fluids, $\omega=\omega(a)$, where $a$ is the scale factor~\cite{Hu2007,Clifton2011,Martinelli2011}.
The best observable to discriminate between MG theories and $\Lambda$CDM is the growth  rate of cosmic 
structures, $f(a)$, defined 
as~\cite{Strauss}
\begin{equation}\label{growth rate}
f(a) \equiv \dfrac{d\ln \delta}{d\ln a}\,,
\end{equation}
where $\delta$ is the matter density contrast. 
The growth rate $f = f(a)$ has indeed the potential to constrain models based on MG theories and $\Lambda$CDM DE (based on GR) 
from the measure of the growth index, $\gamma$, when one parametrizes $f$ as~\cite{Linder}
\begin{equation}\label{growth rate parametrize}
f(a) = \Omega_{m}(a)^{\gamma} \, .
\end{equation}
In the $\Lambda$CDM concordance model,  $\gamma \simeq 0.55$, and for some MG  models $\gamma$ evolves with time: $\gamma \simeq 0.41 -  0.21z$~\cite{Basilakos12}. 
In fact, changing the gravitational theory will affect the way how matter clumps at all scales, beyond the expansion of the homogeneous universe. 
As such, alternative cosmological models based on MG theories make very different predictions for matter clustering and evolution.

Direct measurements of $f(a)$, or equivalently of $f(z)$, where the scale factor and the redshift are related by $a=1/(1+z)$, 
are difficult to obtain. 
Instead the data available comes from the  measurements of the galaxy redshift space distortions in the form of a product of two quantities 
evolving with time: $[f\sigma_8](z) \equiv f(z) \sigma_8(z)$, where $\sigma_8$ is the  root-mean-square linear fluctuation of the matter 
distribution at the scale  of 8 Mpc$/h$, $h$ the dimensionless Hubble parameter, where $H_0 = 100 \,h$ km/s/Mpc. 
Notice that the letter $f$ refers to the growth rate of cosmic structures, that is, it represents the function $f(a)$ or $f(z)$, but it is commonly used in the form $f(R)$ to refer to MG theories. 
In what follows the symbol $f(R)$, as a function of $R$, only refers to MG theories, while the letter $f$ alone or the  function $f(a)$, or $f(z)$, refers to the cosmic growth rate.
In this paper, we use this observational parameter, $[f\sigma_8]$ and the background evolution of the universe, $H$ to explore the parameter space and test a model of $f(R)$ gravity, namely, the Starobinsky $f(R)$ model.

We use a Markov Chain Monte Carlo (MCMC) method to explore the parameter space of the theory with the selected datasets for $H(z)$ and $[f\sigma_8](z)$, 
both individually and jointly. In the end, we further extend our analysis with a joint study of the observables proposed by Linder~\cite{Linder2017} and, 
in the context of Modified Gravity, more recently by Matsumoto~\cite{Matsumoto2020}.

This work is organized as follows. 
A brief description of the Starobinsky cosmological model~\cite{Starobinsky2007} is presented in section~\ref{sec2}. 
In section~\ref{data} we provide details of the two data sets used in the analyses, and in section~\ref{analyses-results} we describe 
the methodology of the analyses performed, show our results, and provide their statistical interpretation. 
The conclusions obtained from these analyses and the final remarks of our work are addressed in  section~\ref{final-remarks}.

\section{Starobinsky f(R) model cosmology} \label{sec2}

\subsection{Background space-time}
\label{sec21}
A general $f(R)$ theory in the metric formalism that is minimally coupled to matter has an action~\cite{DeFelice2010}
\begin{equation}
\label{action_fr}
S = \int d^4 x \sqrt{-g}\left[ \frac{1}{2\kappa}f(R) + \mathcal{L}_m(g_{\mu\nu})   \right],
\end{equation}
where we include in the action 
the matter Lagrangian density,  $\mathcal{L}_m$, and define $\kappa \equiv 8\pi G$ 
(we adopt units with $c = 1$). 
This allows us to use geometrized units more easily when needed.
In this work we define the function $f(R)$ as in the model introduced  by Starobinsky \cite{Starobinsky2007} 
\begin{equation}
\label{fr_st}
f(R) \equiv R + \lambda R_0 \left(\left(1+\frac{R^2}{R_0^2}\right)^{-n} -1\right).
\end{equation}
$R_0$ and $\lambda$ are model parameters related to the observed DE density and a characteristic curvature, respectively~\cite{Chen2019}, 
which can be constrained by the measured cosmological parameters using equation (\ref{cosmparam}). 
In this paper we will analyse the cases $n=1$ and $n=2$, which are extensively studied in the literature \cite{Kopp_2013,Chen2019}. 
While the model with $n=1$ is known to have difficulty in passing solar system tests and reproducing the matter density fluctuations power spectrum \cite{Starobinsky2007,MOTOHASHI_2009}, it is still used in the literature as a prototypical example of the theory, as well as a fit to the data 
and MCMC analysis \cite{Chen2019,Cardone2012,Hough2020}. The parameter range for models which pass the Solar System test is $n\geq 2$, 
which also has the most pronounced effect on the evolution of matter perturbations \cite{Motohashi_2013,Chudaykin_2015}.

The variation with respect to the metric gives rise to the field equations of this model, given by the extended Einstein equations 
\begin{align}
&F(R) R_{\mu\nu} - \frac{1}{2}f(R)g_{\mu\nu} - [\nabla_\mu\nabla_\nu - g_{\mu\nu}\square]F(R) = \kappa T_{\mu\nu},\\
&F(R)R - 2f(R) + 3\,\square f(R) = \kappa T, \label{Field_Equations}
\end{align}
where we define $F(R) \equiv d\/f(R) / d R$, $T \equiv g^{\mu\nu}T_{\mu\nu}$, and the second equation is the trace of the first one, useful to derive the dynamics of the $f(R)$ function, which is more complicated than in GR where one has  an algebraic relation between $R$ and $T$ \cite{Sotiriou2010}.

To model our cosmological space-time 
we adopt the  Friedmann-Lema\^{\i}tre-Robertson-Walker (FLRW) for the background space-time  filled with a perfect fluid,  
with metric and stress-energy tensor 
given by, respectively, 
\begin{align}
&ds^2 = -dt^2 + a^2(t)\left[\frac{dr^2}{1 - Kr^2}  + r^2 (d\theta^2 + \sin^2\theta \,d\phi^2)   \right] \label{FLRW},\\
&T^{\mu\nu} = (\rho + P)u^\mu u^\nu + Pg^{\mu\nu} \label{perfectfluid},
\end{align}
where we take, as usual, $K$ is the scalar curvature of the 3-space, and $\rho$ and $P$, are the density and pressure of the perfect fluid,  respectively. $u^{\mu}$ is the 4-velocity vector of an observer comoving with the fluid; 
$a=a(t)$ is the dimensionless scale factor, with $a(t_0)=1$, $t_0$ is today's cosmic time, and we always use geometrized units, unless otherwise noted.
Because we shall compare our analyses with those of the flat $\Lambda$CDM model, we consider the flat FLRW  space-time, i.e.,  $K=0$. 

One can analyse the dynamical evolution of our cosmological model using equations~\eqref{FLRW} and  \eqref{perfectfluid} in 
equation~\eqref{Field_Equations}, obtaining the modified Friedmann equations 
\begin{align}
\label{modified_friedmann}
&\!H^2 = \frac{\kappa}{3F(R)}\left[ \rho + \frac{RF(R) - f(R)}{2} - 3H\dot{R} F'(R) \right] ,& \\ 
&2\dot{H} + 3H^2 = -\frac{\kappa}{F(R)}
\left[ P + \dot{R}^2F''(R) + 2H\dot{R}F'(R) 
\right. & 
\nonumber \\
& \hspace{2.0cm}+ \, \ddot{R}F'(R) + \frac{1}{2} (f(R) - RF(R)) \Big],& 
\end{align}
where $\dot{\,}$ means derivative with respect to the cosmic time $t$. 
These equations must obey certain constraints in order that our model is stable and correctly reproduces the late-time acceleration of the universe,  without deviating from $\Lambda$CDM at early times  (associated with $R \gg R_0$). An analysis of such constraints on general $f(R)$ theories can be found in \cite{Sotiriou2010}. Following \cite{Starobinsky2007}, it suffices that \eqref{fr_st} should satisfy, for the values $n=1$ and $n=2$
\begin{align}
\label{preconstraints}
    F(R) > 0\,, \quad F'(R) > 0\,, \quad R \geq R_1\,, \quad \lambda > \frac{8\sqrt{3}}{3}\,, \quad  &n=1,\\
    \lambda > \frac{\sqrt{(\sqrt{13}-2)}}{2}\,, \quad &n=2
\end{align}
where $R_1$ is the Ricci curvature of a de Sitter point in the space of solutions. More stringent constraints on the model parameters $n$ and $R_1$, can be 
obtained by likelihood analyses on various data sets for the Starobinsky and other  $f(R)$ models, like the ones performed in ref.~\cite{Cardone2012}, 
whereas more extensive and general analyses of the stability of the model can be found in ref.~\cite{Motohashi_2011_1}.

It will be convenient to write equation~\eqref{fr_st} in terms of the measured cosmological parameters $\Omega_{m0} = \frac{8\pi G \rho_{m0}}{3 H_0^2}$ and $\Omega_{\Lambda0} = \frac{8\pi G \rho_{\Lambda}}{3 H_0^2}$. In this case, one can find the correct way to write the parameters by making the model return to $\Lambda$CDM at high $z$ limit \cite{Starobinsky2007,Chen2019} 
\begin{align}
\label{cosmparam}
f(R) =\,\, & R \,+\, 6\lambda H_0^2(1-\Omega_{m}) & \nonumber \\
& \times \left(\left(1+\frac{R^2}{[6H_0^2(1-\Omega_{m})]^2}\right)^{-n} -1\right) \,.&
\end{align}
This relation allows us to constrain $R_0$ and $\lambda$ using data linked to the observables $\Omega_{m0}$ and $H_0$.

\subsection{Metric  perturbations}
In order to go beyond the background cosmological observables and test our theory against structure growth we need to introduce  metric perturbations.  Following the general perturbative procedure for scalar-tensor theories found in \cite{Tsujikawa2007}, in the usual Newtonian (or comoving) gauge, the perturbed metric is given by 
\begin{equation}
    ds^2 = -(1+2\Psi)\/dt^2 + (1-2\Phi)\,a^2(t)\,\delta_{ij} dx^i dx^j\,,
\end{equation}
where $i,j=1,2,3$, and $\Psi, \Phi$ are the Bardeen potentials which satisfy $\Psi = \Phi$ in the absence of anisotropic stress. 
The physical processes we are interested in, and the cosmological observables associated with them (i.e., 
accelerated cosmic expansion and growth of cosmic structures), are all well within the scale of the sub-horizon approximation $k \gg a^2H^2$. In this case, we can write the evolution equation for the matter density contrast, $\delta$, as 
\begin{equation}
\label{evo_time}
    \ddot{\delta} + 2H\Dot{\delta} + \frac{k^2}{a^2}\Phi = 4\pi G\, \mu(k,a)\,\rho_m\,\delta \,,
\end{equation}
where {$G$} is the Newtonian gravitational constant. 
The $\mu(k,a)$ factor is written as 
\begin{equation}
\mu(k,a) \equiv \frac{1}{8\pi F(R)}\left(\frac{1 + 4\frac{k^2}{a^2 R}\,m}{1 + 3\frac{k^2}{a^2 R}\,m}\right),
\end{equation}
where $m$ is a parameter that quantifies the deviation from the $\Lambda$CDM model 
\begin{equation*}
    m \equiv \frac{R F^\prime(R)}{F(R)} \implies m \vert_{{\footnotesize \Lambda CDM}} = 0 \,.
\end{equation*}
Equation \eqref{evo_time} allow us to analyse how our cosmological model should behave at the limits of very small and very large scales; 
since $R$ is of the order of $H(z)$, we obtain
\begin{align}
    &\lim_{k \gg a^2 H^2} \mu(k) = \frac{4}{3}\,\frac{1}{8\pi F(R)}, \\
    &\lim_{k \ll a^2 H^2} \mu(k) = \frac{1}{8\pi F(R)} \,,
\end{align}
and we observe that, at small scales, the modification of GR gives an extra factor of $\frac{4}{3}$ to the force term of the equation, therefore gravity becomes stronger and cosmic structures grow faster. 
In the opposite limit, at large scales, the equation \eqref{evo_time} is the same as in GR \cite{Tsujikawa2007}, now with the term $1/F(R)$, which arises naturally in a theory that couples to gravity. 
In the language of the Scalar-Tensor theory, this would be the term that couples gravity to the scalar field.

The regime of cosmic structure formation lies between these scales, so that we need the full equation \eqref{evo_time} to account for the physics in this 
regime. However, when the Hubble scale is not large enough in comparison to the perturbations (high redshift), or when we have scales of the order of $H_0$ 
(low redshift), the behavior of the structure formation should have the $\Lambda$CDM model as a limit.

It is useful to rewrite equation \eqref{evo_time} as an ODE on the variable  $N \equiv \ln{a}$, also called the e-fold number, which can be obtained after a simple chain rule derivation 
\begin{equation}
\label{growth_fr}
\delta'' + \left(2 + \frac{H'}{H}\right)\delta' - \frac{3}{2}\Omega_m(a)\,
\mu(k,a)\,\delta = 0 \,,
\end{equation}
where $'$ means derivative with respect to $N$, and $\Omega_{m}(a)$ is the density parameter of (dark + baryonic) matter 
as a function of the scale factor $a$, defined above. 
This is the equation to be solved in order to find the linear growth of our model, using the definition of the growth rate 
of cosmic structures function $f(a)$ given in equation~\eqref{growth rate} in the same way as in GR.


%
\section{Observational Data}\label{data}

There are in the literature several compilations of $H(z)$ and $[f \sigma_8](z)$ observational data. In this section, 
we are going to present the datasets we are considering to impose observational constraints in the Starobinsky model.
\subsection{H(z) data}
The history of the expansion of the universe is probed by the observational Hubble parameter $H(z)$. It can be measured by several independent methodologies, most of them, based on distance measurements of galactic objects such as supernovae and quasars. 
However, it is known that in some approaches the cosmological distance measurements depend on a fiducial model, which makes the use of these data problematic when the objective is to constrain the free parameters of cosmological models in alternative scenarios.  

In this work, since we are considering a gravity model of type $f(R)$, we are going to use in our analyses, the $H(z)$ measurements obtained by the differential age technique, which is independent of the fiducial model. 
This technique, also known as cosmic chronometers (CC), was proposed by Jimenez and Loeb in \cite{Jimenez_2002}.

The basic idea of the CC data consists of the  spectroscopic determination of the age difference between two passively evolving early-type galaxies. The assumption of  old galaxies to realize the age difference measurements is important to assure that the galaxies were formed at the same time, although are localized in slightly different redshift.

\begin{figure}[h!]
\resizebox{0.48\textwidth}{!}{%
  \includegraphics{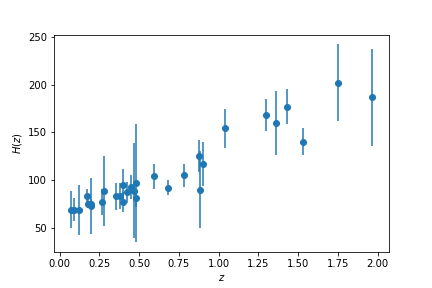}
}
\caption{The CC Hubble parameter data, $H(z)$, displayed in Table \ref{table_1}.}
\label{figure_1}       
\end{figure}

The Hubble parameter is directly related to the measured quantity $dz/dt$ 
by 
\begin{equation}
H(z) = -\frac{dz}{dt}\frac{1}{(1+z)} \, .
\end{equation}
In this work we use the compilation of $31$ measurements of $H(z)$, covering the 
range $0.07 < z < 1.965$~\cite{Yang_2020}.  The data is shown in Table \ref{table_1} and in Fig. \ref{figure_1}.

\begin{table} [ht!]
\centering
\caption{The $31$ Hubble parameter data points, $H(z)$, and their respective $1\,\sigma$ errors  $\sigma_H(z)$ from the CC data~\cite{Yang_2020}. 
The units for both $H(z), \sigma_H(z)$ are km/s/Mpc.}
\label{table_1}   
\begin{tabular}{|l|l|l|}
\hline\noalign{\smallskip}
$z$    & $H(z)$ & $\sigma_H(z)$\\  
\noalign{\smallskip}\hline\noalign{\smallskip}
0.07   & 69.0 & 19.6 \\  
0.09   & 69.0 & 12.0 \\   
0.12   & 68.6 & 26.2 \\    
0.17   & 83.0 & 8.0  \\     
0.179  & 75.0 & 4.0  \\    
0.199  & 75.0 & 5.0  \\    
0.2    & 72.9 & 29.6 \\     
0.27   & 77.0 & 14.0 \\     
0.28   & 88.8 & 36.6 \\
0.352  & 83.0 & 14.0 \\
0.3802 & 83.0 & 13.5 \\
0.4    & 95.0 & 17.0 \\
0.4004 & 77.0 & 10.2 \\
0.4247 & 87.1 & 11.2 \\
0.4497 & 92.8 & 12.9 \\
0.47   & 89.0  & 49.6 \\
0.4783 & 80.9  & 9.0  \\
0.48   & 97.0  & 62.0 \\
0.593  & 104.0 & 13.0 \\
0.68   & 92.0  & 8.0  \\
0.781  & 105.0 & 12.0 \\
0.875  & 125.0 & 17.0 \\
0.88   & 90.0  & 40.0 \\
0.9    & 117.0 & 23.0 \\
1.037  & 154.0 & 20.0 \\
1.3    & 168.0 & 17.0 \\
1.363  & 160.0 & 33.6 \\
1.43   & 177.0 & 18.0 \\
1.53   & 140.0 & 14.0 \\
1.75   & 202.0 & 40.0 \\
1.965  & 186.5 & 50.4\\
\noalign{\smallskip}\hline
\end{tabular}
\end{table}

\subsection{$[f \sigma_8](z)$ data} \label{Sec1}

\begin{figure}[h!]
\resizebox{0.48\textwidth}{!}{%
  \includegraphics{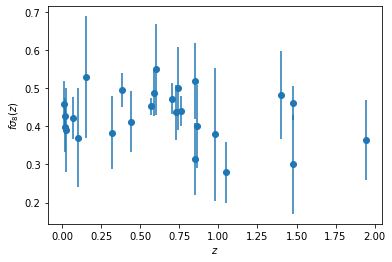}
}
\caption{The $[f\sigma_8](z)$ data displayed in Table \ref{table_2}.}
\label{figure_2}       
\end{figure}

Precise measurements of $[f \sigma_{8}](z)$ can be done using the Redshift Space Distortions (RSD) approach, 
by studying the peculiar velocities caused by local gravitational potentials that introduce distortions in the 
two-point correlation function of cosmic  objects~\cite{1987MNRAS.227....1K}. 
In fact, the calculation of the anisotropic  two-point correlation function, $\xi(s, \mu)$~\cite{Hamilton1992ApJ}, 
allows us to measure $f\sigma_{8}$, where $\sigma_{8}$ is
the root-mean-square linear fluctuation of the matter distribution at the scale of 8 Mpc$/h$ 
(for other approaches to study matter clustering see, e.g., \cite{Avila19,Marques20,deCarvalho18,deCarvalho20}). 
The literature reports diverse compilations of measurements of $[f\sigma_8](z)$ (see, e.g.,~\cite{Nunes:2015xsa,Alam21}) which we update here. 
Our compilation, shown in Table \ref{table_2} and in Fig. \ref{figure_2}, takes into account: 
\begin{enumerate}
    \item we consider $[f\sigma_8](z)$ data obtained from disjoint or uncorrelated redshift bins when the measurements concern the same cosmological tracer, and data from possibly correlated redshift bins when different cosmological tracers were analyzed; 
    \item we consider the latest measurement of $[f\sigma_8](z)$ when the same survey 
collaboration performed two or more measurements corresponding to several data releases. 
\end{enumerate}

\linespread{1.2}
\begin{table*}
\caption{Updated compilation of 26 $[f\sigma_8](z)$ data with their  respective $1\sigma$ errors, $\sigma_{f\sigma_8}$.
}
\centering
\label{table_2}
\begin{tabular}{|c|c|c|c|c|}
\hline\noalign{\smallskip}
Survey & $z$ & 
$[f \sigma_8] \pm \sigma_{f\sigma_8}$ & Reference & Cosmological tracer \\
\noalign{\smallskip}\hline\noalign{\smallskip}
ALFALFA    & 0.013 & $0.46 \pm 0.06$ & \cite{Avila21a} & HI line sources \\
SNIa+IRAS    & 0.02 & $0.398 \pm 0.065$ & \cite{Turnbull} & galaxies+SNIa \\
SNIa+6dFGS & 0.02 & $0.428 \pm 0.0465$ & \cite{Huterer} & galaxies+SNIa \\
6dFGS & 0.025  & $0.39 \pm 0.11$ & \cite{Achitouv17} & voids \\
6dFGS & 0.067 & $0.423 \pm 0.055$ & \cite{Beutler} & galaxies \\
SDSS-veloc & 0.10  & $0.370 \pm 0.130$ & \cite{Feix} & galaxies \\
SDSS-IV & 0.15  & $0.53  \pm 0.16$ & \cite{Alam21} &
MGS \\
BOSS-LOWZ   & 0.32 & $0.384 \pm 0.095$ & \cite{Sanchez} & LOWZ and CMASS samples \\
SDSS-IV & 0.38  & $0.500 \pm 0.047$ & \cite{Alam21} & BOSS galaxy sample \\
WiggleZ   & 0.44  & $0.413 \pm 0.080$ & \cite{Blake12} & ELG \\
BOSS-CMASS & 0.57  & $0.453 \pm 0.022$ & \cite{Nadathur} & voids+galaxies \\
SDSS-CMASS & 0.59 & $0.488 \pm 0.060$ & \cite{Chuang16} & CMASS galaxy sample \\
VIPERS PDR-2 & 0.60 & $0.550 \pm 0.120$ & \cite{Pezzotta17} & galaxies \\
SDSS-IV & 0.70  & $0.448 \pm 0.043$ &  \cite{Alam21} & eBOSS DR16 LRG \\
WiggleZ  & 0.73  & $0.437 \pm 0.072$ & \cite{Blake12} & ELG \\
SDSS-IV & 0.74  & $0.50 \pm 0.11$ &  \cite{Aubert20} & eBOSS DR16 Voids \\
Vipers v7 & 0.76  & $0.440 \pm 0.040$ &  \cite{Wilson} & galaxies \\
SDSS-IV & 0.85  & $0.52 \pm 0.10$ &  \cite{Aubert20} & eBOSS DR16 Voids \\
SDSS-IV & 0.85  & $0.315 \pm 0.095$ &  \cite{Alam21} & eBOSS DR16 ELG \\
VIPERS PDR-2 & 0.86 & $0.400 \pm 0.110$ &  \cite{Pezzotta17} & galaxies \\
SDSS-IV & 0.978 & $0.379 \pm 0.176$ &  \cite{Zhao} & eBOSS DR14 Quasar \\
Vipers v7 & 1.05  & $0.280 \pm 0.080$ &  \cite{Wilson} & galaxies \\
FastSound & 1.40 & $0.482 \pm 0.116$ &  \cite{Okumura} & ELG \\
SDSS-IV & 1.48  & $0.30 \pm 0.13$ &  \cite{Aubert20} & eBOSS DR16 Voids \\
SDSS-IV & 1.48   & $0.462 \pm 0.045$ &  \cite{Alam21} & eBOSS DR16 Quasar \\
SDSS-IV & 1.944 & $0.364 \pm 0.106$ &  \cite{Zhao} & eBOSS DR14 Quasar \\
\noalign{\smallskip}\hline
\end{tabular}
\end{table*}
\linespread{1.0}

\subsection{Joint data}
The binned data given in Table~\ref{table_3} was obtained  binning simultaneously the $[f\sigma_8](z)$ and $H(z)$ 
(from the Table \ref{table_2} and the Table \ref{table_1}, respectively) in $5$ redshift bins: 
$(0.0,0.30]$, \,$(0.30,0.60]$,  \,$(0.60,0.85]$, \,$(0.85,1.4]$, \,$(1.4,2.0]$, 
with mean redshifts: $\bar{z} = \,0.15, \,0.45, \,0.725, \,1.125, \,1.7$. 
The number of data pairs $(f\sigma_8, H)$ in each bin were: 
$(7,9)$, $(6,10)$, $(6,2)$, $(4,6)$, 
$(3,4)$, respectively. 
The values in these $5$ bins for $[f\sigma_8](z)$ and $H(z)$, and their errors, correspond to their  variance weighted means~\cite{Moresco2017}. 
This binned data will be used in the $\chi^2$ joint analysis.


\linespread{1.25}
\begin{table}[h!]
\centering
\caption{Binned data of $[f\sigma_8](z)$ and $H(z)$ obtained by calculating 
the variance weighted mean in each redshift bin (see the text for details).} 
\vspace{0.2cm}
\begin{tabular}{|c|c|c|} 
\hline\noalign{\smallskip}
$\bar{z}$ & $[f \sigma_8](z)$ & $H(z)$  \\
\noalign{\smallskip}\hline\noalign{\smallskip}
0.15   & $0.4284 \pm 0.0338$ & $75.367 \pm 5.741$ \\  
0.45   & $0.4647 \pm 0.0288$ & $88.880 \pm 6.717$ \\   
0.725 & $0.4433 \pm 0.0313$  & $98.500 \pm 7.071$ \\    
1.125 & $0.3852 \pm 0.0602$  & $135.667 \pm 10.247$ \\
1.70   & $0.3753 \pm 0.0541$ & $176.375 \pm 15.300$ \\
\noalign{\smallskip}\hline
\end{tabular}\label{table_3}
\end{table}
\linespread{1.}

\section{Analyses and Results} \label{analyses-results}

In this section we describe the parameter space analyses of the $f(R)$ model given in equation~\eqref{fr_st}, using the  observational data described in the previous section \ref{data}. 
To obtain the theoretical predictions for $H(z)$ and $f(z)$ from the chosen model,  equation~\eqref{action_fr}, we integrate equations \eqref{modified_friedmann} and \eqref{growth_fr} numerically and use their results in the assembling of the likelihood function. 
This gives us our statistical analyses.  

The standard Bayesian inference will be considered for the  parameter estimation. To investigate how appropriate the data is to constrain the parameter space of the model, they will be considered separately and later combined in a joint analysis. 


\subsection{$H(z)$ and $[f\sigma_8](z)$ observational constraints} \label{MCMC-analyses}

Recently, the tension between different  estimates of the Hubble constant has drawn the attention of cosmologists. 
The local  determination of the Hubble constant from SHO-ES 
collaboration \cite{Riess:2019cxk} is  $H_0^{\mbox{\tiny SHOES}}= 74.03 \pm 1.42 $ km/s/Mpc  while the value inferred by the Planck collaboration \cite{Planck2018}, from the Cosmic Microwave Background (CMB) analysis in a flat $\Lambda$CDM framework is $H_0^{\mbox{\tiny Planck}} = 67.36 \pm 0.54$ km/s/Mpc from the last Planck collaboration  results~\cite{Planck2018}. 

The divergence in $H_0$ measurements is the main reason to look for alternative approaches to perform analyses that are independent of the $H_0$ value \cite{NB20}. 
In our analyses, we shall consider $H(z)/H_0$ instead of $H(z)$ data, and for this we use 
$H_0 = H_0^{\mbox{\tiny  Planck}}$~\cite{Planck2018}. 



For the model given in equation~(\eqref{fr_st}) and $H(z)$ data, the Likelihood function is given by 
\begin{equation}
\mathcal{L}(z|\theta) \propto -\frac{1}{2}\sum _{i}^{N_H} \,
\frac{( H^i_{th}(z|\theta) - H^i_{obs}(z))^2}{\sigma_i ^2} \, ,
\end{equation}
where $N_H$ is the number of points in the dataset, $\theta={(\Omega_{m0},\lambda )}$ are the 
free parameters of the model, $H^i_{th}(z|\theta)$  is the theoretical Hubble parameter at redshift 
$z_i$, $H^i_{obs}(z)$ and $\sigma$  are the observation error values of the Hubble parameter given in table \ref{table_1}.

The theoretical Hubble parameter depends on the current Hubble parameter $H_0$.
To eliminate this dependency, we will follow the approach introduced by \cite{Anagnostopoulos}, 
that consists of marginalization of the likelihood over $H_0$ parameters 
\begin{equation}
\mathcal{L}= \Gamma -\frac{B^2}{A} + \ln{A} - 2\ln{\left[1+\textrm{erf}  
\left(\frac{B}{\sqrt{2A}}\right)\right]} \, ,
\end{equation}
where, by the definition of $E(z) \equiv H_{th} / H_0$, 
\begin{equation}
A = \sum^N_{i=1} \frac{E^2(z_i)}{\sigma_i^2} \, ,
\end{equation}
\begin{equation}
B = \sum^N_{i=1} \frac{E(z_i)H_{obs}(z_i)}{\sigma_i^2} \, ,
\end{equation}
and
\begin{equation}
\Gamma = \sum^N_{i=1} \frac{H_{obs}^2(z_i)}{\sigma_i^2} \, .
\end{equation}

Similarly, the Likelihood function for $[f\sigma_8](z)$ data is given by:
\begin{equation}
\mathcal{L}(z|\theta) \propto -\frac{1}{2}\sum _{i}^{N_f} \,
\frac{( f\sigma_8^{i_{th}}(z|\theta) - f\sigma_8^{i_{obs}}(z))^2}{\sigma_i ^2} \, ,
\end{equation}
where we have the same parametric space $\theta={(\Omega_{m0},\lambda )}$, $ f\sigma_8^{i_{th}}(z|\theta)$  is the theoretical growth function given by the model with parameters $\theta$ at redshift 
$z_i$; $f^i_{obs}(z)$ and $\sigma$ are the observed values, given by the data in Table \ref{table_2}; $N_f$ is the size of the $f\sigma_8$ dataset.
The $f\sigma_8$ function  does not have $H_0$ as a free parameter and therefore does not need to be marginalized. The parametric space of both observables $H(z)$ and $[f\sigma_8](z)$ are the same and the joint likelihood function is given by the product of the individual
likelihoods according to

\begin{equation}
\mathcal{L} = \mathcal{L}_{H}\times \mathcal{L}_{f\sigma_8}.
\end{equation}


The Bayesian inference of the parameters is obtained through the expected values of the posterior density $p(\theta|z)$ and since the posteriori distribution is unknown for the model, i.e., it can not be approximated by a normal or gaussian distribution, we are considering the Bayes Theorem that establishes
\begin{equation}
   p(\theta|z) = \mathcal{L}(z|\theta) \times \Pi(\theta)
\end{equation}
where $\Pi(\theta)$ is the prior.

The parametric space of the variables of the model will be explored following the methodology of Markov chain Monte Carlo (MCMC) and the Metropolis–Hastings algorithm to generate a set of samples from a posteriori distribution. To implement the MCMC, we are using a Python open-source code, {\fontfamily{cmtt}\selectfont emcee}  \cite{Foreman_Mackey_2013}

The posteriori distribution was generated for the following intervals in the $n=1$ case, with flat distributions: $\lambda^{-1} \in[0.001, 1.8]$ and $\Omega_{m0} \in [0.2, 0.6]$ for the $H(z)$ data set; $\lambda^{-1} \in [0.001, 2.0]$ and $\Omega_{m0} \in [0.15, 0.3]$ for the $[f{\sigma 8}](z)$ data set; and finally, $\lambda^{-1} \in [0.001, 2.0]$ and  $\Omega_{m0} \in [0.2, 0.4]$ for the combined data set.
For the $n=2$ case we use a flat prior in all the MCMC runs, as well, we consider the same intervals $\lambda^{-1} \in [0.1, 2]$, $\Omega_{m0} \in [0.15, 0.4]$. 

\begin{figure}[h]
\centering
  \includegraphics[width=8cm]{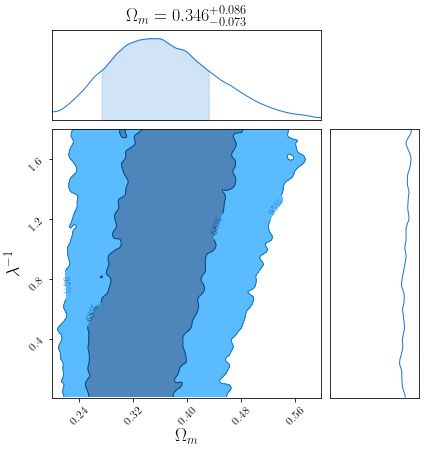}
\caption{MCMC simulations for the $f(R)$ model with $n=1$, considering $H(z)$ data and the flat priors: 
$\lambda^{-1} = [0.001, 1.8]$ and $\Omega_{m0} = [0.2, 0.6]$.}.
 \label{figure_3}
\end{figure}

\begin{figure}[h]
\centering
  \includegraphics[width=8cm]{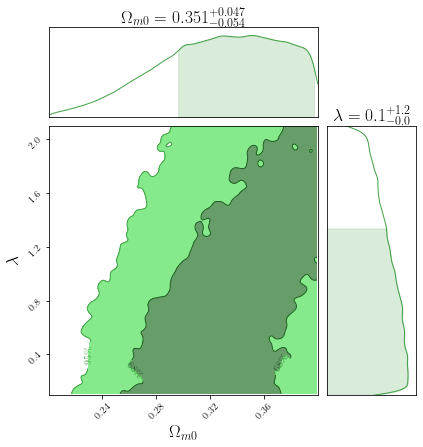}
\caption{MCMC simulations for the $f(R)$ model with $n=2$, considering $H(z)$ data and the flat priors: 
$\lambda^{-1} \in [0.1, 2]$ and $\Omega_{m0} \in [0.15, 0.4]$.}.
 \label{figure_32}
\end{figure}

\begin{figure}[h]
\centering
  \includegraphics[width=8cm]{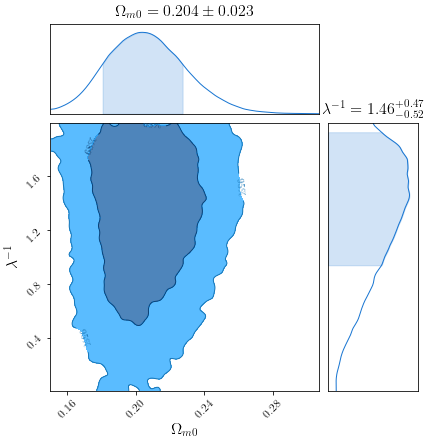}
\caption{MCMC simulations for the $f(R)$  model with $n=1$, considering $[f\sigma_8](z)$ data and the flat priors: 
$\lambda^{-1} \in [0.001, 2.0]$ and $\Omega_{m0} \in [0.15, 0.3]$.}
 \label{figure_4}
\end{figure}

\begin{figure}[h]
\centering
  \includegraphics[width=8cm]{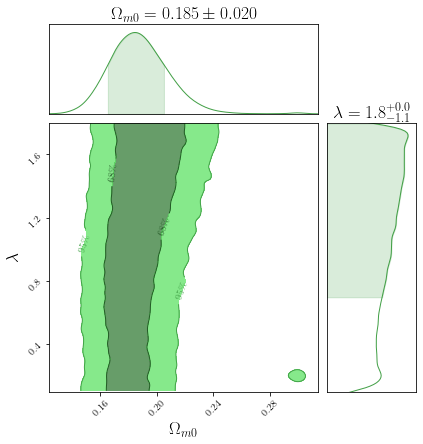}
\caption{MCMC simulations for the $f(R)$  model with $n=2$, considering $[f\sigma_8](z)$ data and the flat priors: 
$\lambda^{-1} \in [0.1, 2.0]$ and $\Omega_{m0} \in [0.15, 0.4]$.}
 \label{figure_42}
\end{figure}

\begin{figure}[h]
\centering
  \includegraphics[width=8cm]{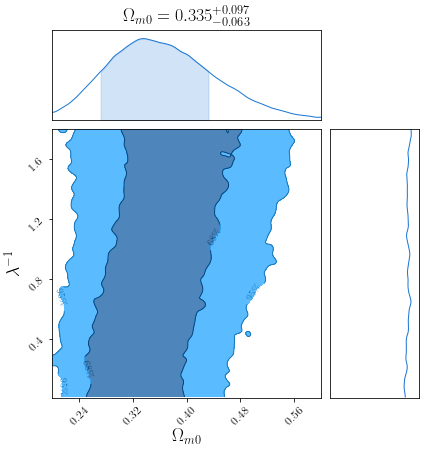}
\caption{MCMC simulations for the $f(R)$ model with $n=1$, considering joint analyses of the  $H(z)$ and $[f\sigma_8](z)$ data and the flat priors: 
$\lambda^{-1} = [0.001, 2.0]$ and $\Omega_{m0} = [0.2, 0.4]$.}
 \label{figure_5}
\end{figure}

\begin{figure}[h]
\centering
  \includegraphics[width=8cm]{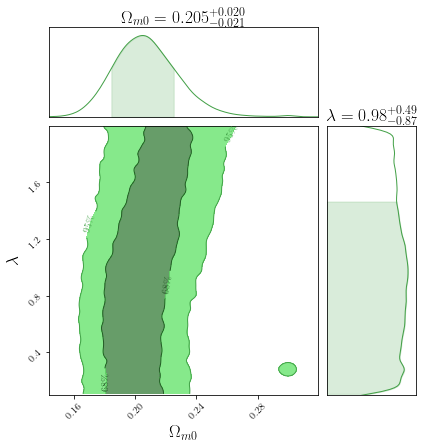}
\caption{MCMC simulations for the $f(R)$ model with $n=2$, considering joint analyses of the  $H(z)$ and $[f\sigma_8](z)$ data and the flat priors: 
$\lambda^{-1} = [0.1, 2.0]$ and $\Omega_{m0} = [0.15, 0.4]$.}
 \label{figure_52}
\end{figure}

We summarize our results in Table \ref{table_4}, and the parameter space contours obtained can be seen in blue in figures \ref{figure_3}, \ref{figure_4}, and \ref{figure_5} for the $n=1$ case; and in green in figures \ref{figure_32}, \ref{figure_42}, and \ref{figure_52} for the $n=2$ case.

\linespread{1.4}
\begin{table}[h!]
\centering
\begin{tabular}{|c|l|l|l|}
\hline
Data & Model &\,\,$\Omega_{m0}$ & \,\,\,$\lambda^{-1}$ \\
\hline
$H$ & n=1 & $0.346^{+0.086}_{-0.073}$ & $1.1936$ (unconstrained)   \\
$f\sigma_8$ & n=1 & $ 0.204{\pm} 0.023 $ & $1.460^{+0.47}_{-0.52}$ \\
$H +f\sigma_8$ & n=1 & $0.335^{+0.097}_{-0.063}$ & $1.761$ (unconstrained)  \\
$H$ & n=2 & $0.351^{+0.049}_{-0.051}$ & $0.38^{+0.98}_{-0.28}$  \\
$f\sigma_8$ & n=2 & $ 0.185{\pm} 0.020 $ & $1.8^{+0.00}_{-1.1}$ \\
$H +f\sigma_8$ & n=2 & $ 0.205{\pm} 0.020 $ & $0.99^{+0.72}_{-0.65}$  \\
\hline
\end{tabular}
\linespread{.9}
\caption{Results of our likelihood analyses for the cosmological parameters and their  uncertainties.}
\label{table_4}   
\end{table}

\bigskip


\subsection{$[f \sigma_8 - H]$ diagram analyses}

Recently, Linder \cite{Linder2017} and Matsumoto et al. \cite{Matsumoto2020} have proposed a joint analysis of the $H(z)$ and  $[f\sigma_8](z)$ cosmological observables as a way to break the degeneracy of DE  and MG models, also allowing to 
recognize the redshift regime where the tested model most affects the growth of cosmic structures.

In Moresco et al. \cite{Moresco2017}, the authors used a joint statistical analysis of both these observables and applied the method of 
Linder \cite{Linder2017} to analyze individually the parameters of $\Lambda$CDM + $\Sigma m_\nu$ and  $w$CDM (for cosmological parameter 
analyses with massive neutrinos see, e.g.,~\cite{Marques19}). 
The authors obtained $1\sigma$ constraints on the parameters from growth structure data at low redshifts ($z < 2$) and the last Planck (2018) 
data release. 
Of particular interest was the degeneracy between models with massive neutrinos and a modified growth parameter $\gamma$, according to this analysis the models investigated provide a better fit than the flat $\Lambda$CDM model, as constrained by the 
Planck mission \cite{Planck2016}. 
The authors then simulated data points in the $f\sigma_8 - H$ plane to distinguish between the models that better fit the data, and found that 
they could be distinguished with high statistical significance using the $f\sigma_8 - H$ plane.
This shows that the joint analysis proposed by Linder \cite{Linder2017} and Matsumoto et al. \cite{Matsumoto2020}
can be contrasted with data to distinguish models that are degenerate in the cosmological parameters fits. 
In particular, we are interested in study how the Starobinsky $f(R)$ model (see the section \ref{sec21}) can be constrained by using 
$[f\sigma_8](z)$ and $H(z)$ data in the diagram proposed by the aforementioned authors.


\linespread{1.}
\begin{figure}[ht!]
\resizebox{0.48\textwidth}{!}{%
\includegraphics{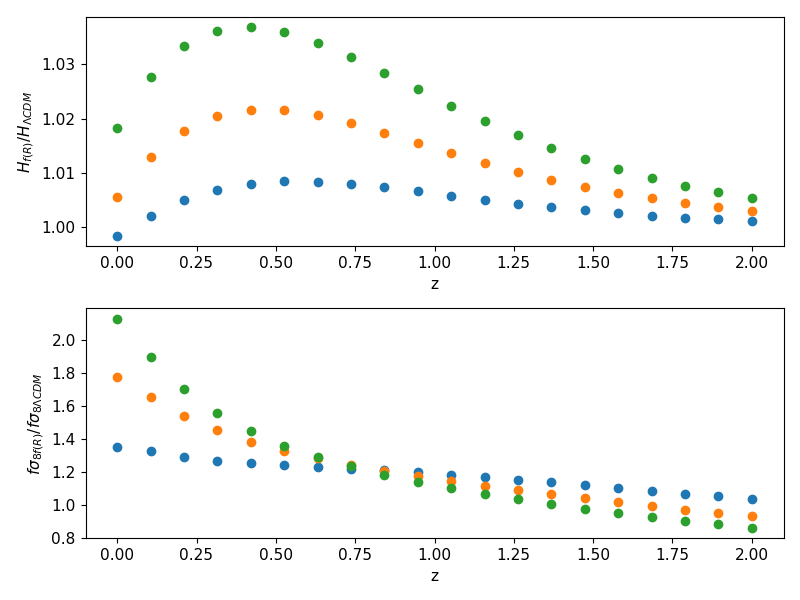}
}
 \caption{Comparison of the evolution of the ratio of observables in the Starobinsnky $f(R)$ models and their $\Lambda$CDM  values. Each point is the theoretical prediction for a given redshift. Different colored curves correspond to different values of the model parameter, $ \lambda^{-1} = 0.3, 0.7, 1)$, with $\Omega_{m0} = 0.3$ and $n=1$}
  \label{figure_6}
\end{figure}

One can build an $f(R)$ modified gravity model that mimics the background evolution of $\Lambda$CDM, like the one presented in section \ref{sec21} and the well-known Hu-Sawicki model \cite{Hu2007}. 
Both models have a similar cosmological evolution, whereas the growth of structures is quite different to the $\Lambda$CDM model, owing to the scale dependence of the solution $\delta (k, a)$ of equation \eqref{growth_fr}. 
In the refs. \cite{Linder2017,Matsumoto2020} it is shown how the $f\sigma_8-H$ plot can help distinguish between MG and $\Lambda$CDM-type models 
(i.e., $\Lambda$CDM or $w$CDM or $w_0 w_a$CDM models with different parameters). 
To perform a similar analysis for our model, we fix the scale of our perturbations at $0.1 h$/Mpc, around the scale where measurements and 
homogeneity assumptions in the $\Lambda$CDM model are made \cite{Avila21a,Avila21b} and a scale where linear perturbation theory is valid in $f(R)$ theories and nonlinear effects can be 
disregarded \cite{Koivisto2006}.

\begin{figure}[ht!]
\resizebox{0.5\textwidth}{!}{%
  \includegraphics{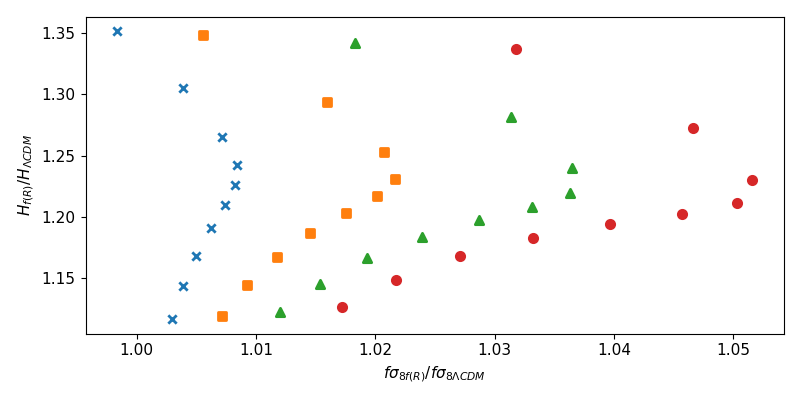}
}
 \caption{Curves on the $[f\sigma_8] / [f\sigma_8]_{\Lambda CDM} \times H/H_{\Lambda CDM}$ plane for different values of the $\lambda$ parameter, for the $n=1$ case. The curves, while having a similar profile, have different curvatures while also having different start and end points. From left to right we have $\lambda^{-1} = 0.3, 0.5, 0.7, 1$.}
    \label{figure_7}
\end{figure}

In figure \ref{figure_6}, we see the separate plot of the ratio of $H(z)$ and $[f\sigma_8] / [f\sigma_{8\Lambda CDM}]$ 
for our Starobinsky $f(R)$ model and their $\Lambda$CDM values, respectively. 
For $H / H_{\Lambda CDM}$ the curves show similarity, also requiring a high amount of precision to distinguish the $< 2\%$ 
absolute difference of the values, whereas for $[f\sigma_8](z)$  there is a high degeneracy in some redshift intervals 
between the same model with different parameters. 
This degeneracy makes it difficult to use data from this redshift interval to constrain the true $\lambda$ parameter of 
the model, while also, again, requiring a higher degree of precision to distinguish between the curves in this interval. 
For higher values of redshift, the degeneracy between the $f\sigma_8$ curves worsens, as they need to converge to the 
$\Lambda$CDM value by construction. Thus the redshift interval that allows us to constrain the true parameter of the model 
with good precision and low degeneracy between the curves is highly dependent on our ability to measure with high precision the local $H(z)$ history and the value of the 
model parameter. 
Using the $f\sigma_8 / f\sigma_{8\Lambda CDM} \times H/H_{\Lambda CDM}$ joint analysis, devised by \cite{Matsumoto2020} and the similar 
$H(z)/H_0 \times [f\sigma_8](z)$ from \cite{Linder2017} to plot the curves for our model, we see that the degeneracy between the different values of 
the parameters is manifestly broken considering the  whole parameters space, as shown in figure \ref{figure_7}.

\begin{figure}
    \includegraphics[width=8cm]{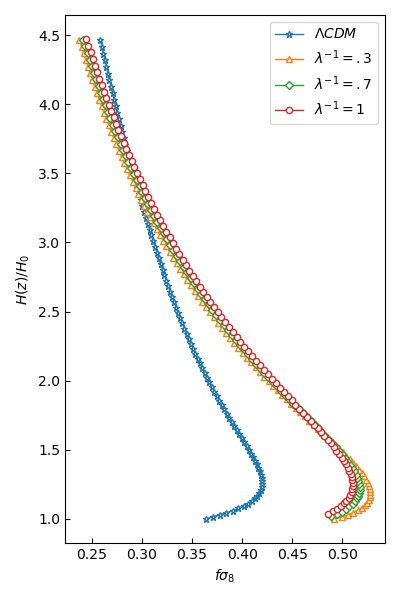}
    \caption{$[f\sigma_8](z) \times H(z)/H_0$ curves for the models 
    $\Lambda$CDM and Starobinsky  $f(R)$ with different values for the parameter $\lambda$ and $n=1$.}
    \label{figure_8}
\end{figure}

With the degeneracy broken, the only limiting factor in distinguishing the curves in the parameter space is the precision 
and the confidence interval of these measurements. 
Figure \ref{figure_8} shows how the parameter space can be further distinguished, including the $\Lambda$CDM model. 
While the expansion history of our (and general) $f(R)$ models follows the $\Lambda$CDM model closely enough that it can't be distinguished from the standard model at the background level,  when one conjoins just the $[f\sigma_8](z)$ evolution with the $H(z)$ expansion of the models, the degeneracy is clarified. 
Although there is still a degree of degeneracy between the curves, with the $H(z)/H_0$ data spanning a large interval, the difficulty in distinguishing 
between models lies, once again, in the precision of the available data. 

In \cite{Linder2017} the author gives an in depth analysis of the relevant redshift range to better distinguish dark energy 
models using this joint analysis of observables, also giving a preview of what future surveys will be able to give precise 
enough measurements to further break the degeneracy between models. 
An extensive analysis using similar methods with other MG models can be found in \cite{Matsumoto2020}. 
In \cite{Moresco2017} a similar analysis is done considering parameter extensions of the concordance $\Lambda$CDM model, with 
a likelihood approach.
Here we'll follow \cite{Moresco2017} in using the $f\sigma_8 - H$ diagram to further distinguish between the parameters of 
our model.

\subsubsection{Goodness of fit between models}

We have produced a set of 5 binned data pairs, $(f\sigma_8(z)$, $H(z))$, 
for an equal number  of redshift bins (see table~\ref{table_3}); they were obtained by calculating the variance weighted mean in each redshift bin. 
We use this data to test  goodness of fit  procedure, {\em via} a $\chi^2$ methodology, to check which set of parameters best fits the joint data, obtained from the MCMC runs in \ref{MCMC-analyses}.
This analysis complements our likelihood analysis where we obtained best fit parameters from the three different sets of data described in \ref{data}. Here, we use the binned data to check which of the pair of parameters obtained from the exploration of parameter space best fits this joint data.
Hereafter we'll call each of the three different best fits obtained from the MCMC run a "model"; the model with the parameters given by the run on just the $H(z)$ data, the one from just $f\sigma_8(z)$ data and the one from the run on both datasets, referring to table \ref{table_4}.

\begin{figure}[h]
\includegraphics[width =  9cm]{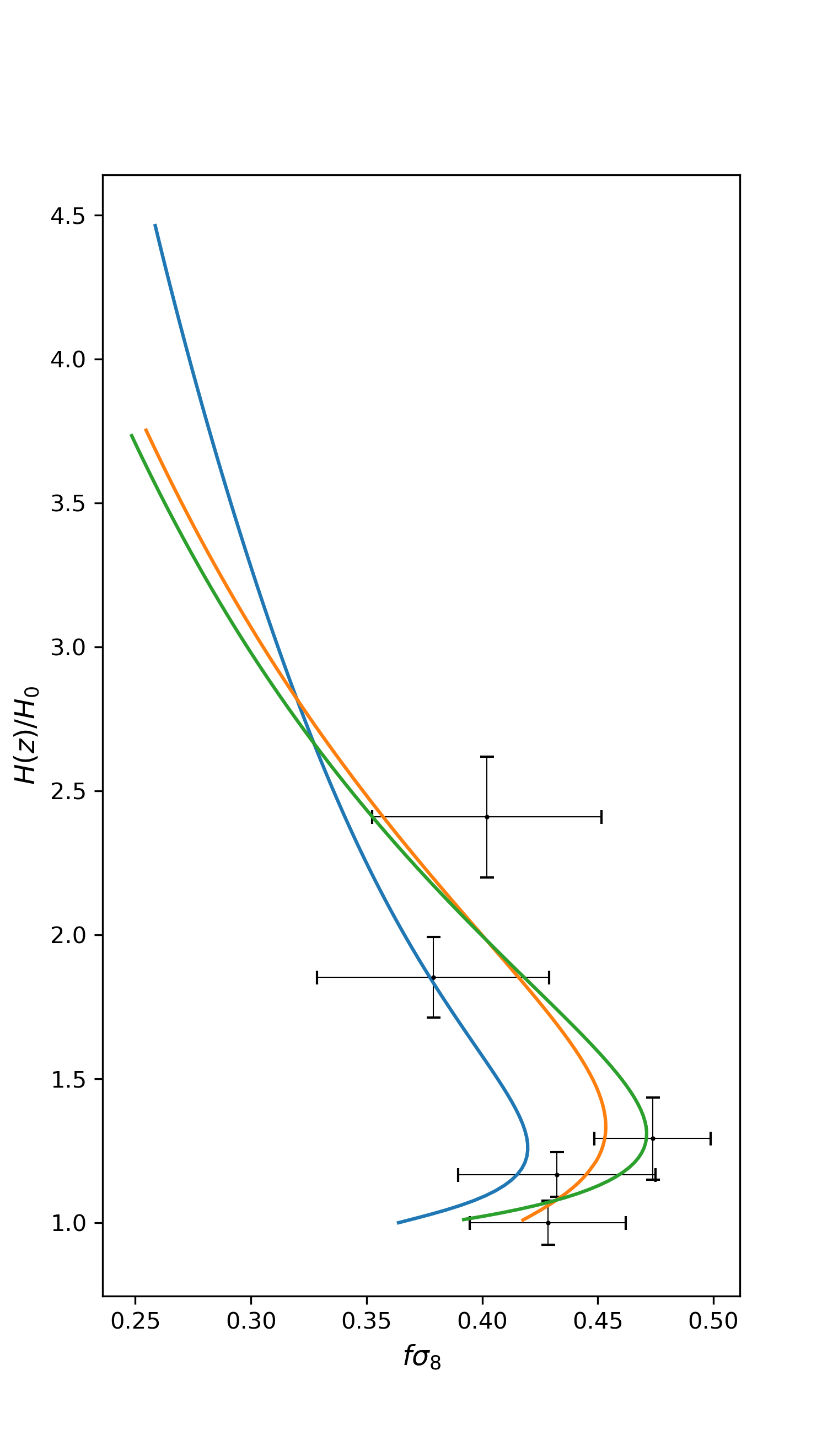}
\caption{Goodness of fit test for binned data, from table \ref{table_3}. The blue curve corresponds to $\Lambda$CDM, the orange one 
to the Starobinsky $n=1$ model with the best-fit parameter $\lambda$, and the green one with the best-fit parameter for $n=2$, both found from the minimization of  eq.~\eqref{chi2_averaged} 
in the MCMC analyses.}
\label{figure_9}
\end{figure}

We devised a simple 2D minimum squared weighted deviation (also a type of $\chi^2$), using the binned data from table \ref{table_3}. 
If, for each data point, we have the individual error on each variable, we have
\begin{itemize}
\item 
$\sigma_{f\sigma_8} \equiv $
$[f\sigma_8](z)$ error,
\item 
$\sigma_{H} \equiv H(z)$  error,
\end{itemize}
and since the data for both observables are independent measurements, i.e., they are not correlated, 
we may write
\begin{equation}
    \sigma^2 \equiv \sigma^2_{\text{joint}} = \sigma^2_{H} + \sigma^2_{f\sigma8},
\end{equation}
as the joint error for both variables  (termed variance in quadrature).

For each model we have a predicted value $(f\sigma_8(z_i)$, $H(z_i))$ at  redshift $z_i$, and a data point $(x_i, y_i)$ in the $f\sigma_8 \times H(z)$ plane. The squared residual between the model prediction $P_i \equiv (f\sigma_8(z_i),H(z_i))$ and the observed data $O_i \equiv (x_i, y_i)$  is given by 
\begin{equation}
(P_i - O_i)^2 =
||(f\sigma_8(z_i) - x_i, H(z_i) - y_i)||^2 \,, 
\end{equation}
where $||\cdot||$ is the distance between the two data points $(f\sigma_8(z_i),H(z_i))$ and $(x_i, y_i)$ in the 
$f\sigma_8 \times H$ plane.

The $\chi^2$ is then defined as the sum of the square of the residuals averaged by the joint error for each data 
\begin{equation}
\label{chi2}
\chi^2 \equiv \sum^n_i \frac{(P_i - O_i)^2}{\sigma_i^2}\,.
\end{equation}
The reduced mean squared weighted deviation is then, for each model analyzed, the ratio between $\chi^2$ 
and the number of degrees of freedom $\nu$, 
$\nu = n - m = 5 - 2 = 3$ for the Starobinsky models, and $\nu = 5 - 1 = 4$ for the flat $\Lambda CDM$ model 
\begin{equation}
\label{chi2_averaged}
\chi^2_\nu = \frac{\chi^2}{\nu} \,.
\end{equation}
In our case the flat $\Lambda$CDM model has one free parameter $\Omega_{\Lambda0}$ (where $\Omega_{m0}$ is constrained by the algebraic relation 
$\Omega_{\Lambda0} + \Omega_{m0} = 1$ ), one less than the Starobinsky models analysed, i.e., $\Omega_{m 0}$ and 
$\lambda^{-1}$. 

Each MCMC parameter space exploration results in a best fit of the model via the likelihood minimization performed by the MCMC method. 
We performed three sets of MCMC runs for each of the cases $n=1$ and $n=2$, each on a different data set, described in section \ref{data}; here we treat each result as a different model, and test the models against the binned data (see table~\ref{table_3}).

\linespread{1.25}
\begin{table*} \centering
\begin{tabular}{|l|c|c|c|} 
\hline 
\hspace{1.8cm} Model & $\Omega_{m0}$ & $\lambda^{-1}$ & $\chi_\nu^2$ \\
\hline 
Starobinsky $n=1$, $\lambda$ best-fit with $H(z)$                   & 0.335 & 1.76 & 6.26 \\ 
Starobinsky $n=1$, $\lambda$ best-fit with $[f \sigma_{8}](z)$ & 0.204 & 1.46 & 0.99 \\ 
Starobinsky $n=1$, $\lambda$ best-fit with joint data          & 0.337 & 1.19 & 5.014 \\ 
Starobinsky $n=2$, $\lambda$ best-fit with $H(z)$             & 0.351 & 0.38  &  3.63\\ 
Starobinsky $n=2$, $\lambda$ best-fit with $[f \sigma_{8}](z)$ & 0.185  & 1.8  & 1.61\\ 
Starobinsky $n=2$, $\lambda$ best-fit with joint data          &0.205  & 0.99 & 1.02\\ 
$\Lambda$CDM                                             & 0.28 &  --    & 0.69 \\ 
\hline
\end{tabular}
\linespread{1.}
\caption{Goodness of fit test, 
measured by $\chi_\nu^2$ through the analysis of the binned data in  table~\ref{table_3}, 
to compare the behavior of the $\Lambda$CDM and 
3 Starobinsky models (each  with a best-fitted $\lambda$ parameter provided by the MCMC runs considering  3 data sets).}
\label{table_5}
\end{table*}

We then calculate \eqref{chi2_averaged} for each of our models and the minimization of this parameter gives a naive estimate of the model that best fits the data in the $f\sigma_8 \times H$ plane. The results of this naive analysis allow us to differentiate between models with different values of $\lambda$ obtained from the MCMC run, which can be seen in table \ref{table_5}. The model that best fits the data according to this criterium is plotted in figure \ref{figure_9}

From the $\chi^2$ statistics, we see that the $\Lambda$CDM model is, from this joint data set, the one with the smallest $\chi^2_\nu$ value, naively giving an "overfit" to the data; with the $n=2$ case having the best fit with all data sets, in comparison with the  $n=1$ case. This is expected, as per the observational problems mentioned before concerning the $n=1$ model, and from the fact that the parameter $n=1$ is in direct relation to deviations from $\Lambda$CDM; $n=1$ being the integer value greater than $0$ that deviates the most from the concordance model. As for the "overfit" of the $\Lambda$CDM model, it's also important to note that the model has 1 parameter less than the $f(R)$ models, which gives it a lower value from the reduced $\chi^2$ statistics.

While one could say that $\Lambda$CDM fits the joint data better than any version of the  Starobinsky models from the $\chi^2$ test, some observations on the methods and data are in order.
First, it's interesting to note that the $f\sigma_8$ data alone gives a better fit to the joint data than the result from the MCMC run on data from both $[f\sigma_8](z)$ and $H(z)$ data, in the $n=1$ case. This could be the case for diverse reasons; one being that the $H(z)$ is actually worsening the results from the $[f\sigma_8](z)$ data when analyzed together. As expected theoretically, background quantities don't constrain modified gravity models suitably; the MCMC run on the $H(z)$ data alone, found in \ref{figure_3}, was not able to constraint the free parameters analyzing the $[f\sigma_8](z)$ data. In the $n=2$ case, the joint data constrains the $\lambda$ parameter significantly better than both the $H$ and $f\sigma_8$ alone, while also giving a better fit from the $\chi^2$ test in general, as can be seen from figure \ref{figure_9}. This points to a fault with the $n=1$ model once again.

Also, the binned data presented in table~\ref{table_3}, built from the $H(z)$ and $[f\sigma_8](z)$ datasets, gives us 5 data points for goodness of fit analysis. 
As a test on whether the binning choice affects the $\chi^2$ statistics, we considered other binned sets and found similar results. 
The advantage of this approach is that it allows to break of the degeneracy between models with dissimilar $\lambda$ parameter, but due to 
the large errors at high$-z$ the analysis is not as accurate as desirable to clearly discriminate between the $\Lambda$CDM and the  Starobinsky models. This could also explain the "overfitting" of the $\Lambda$CDM model, which would give a $\chi^2$ closer to 1 if the error was better constrained.
In fact, more data points and tighter error constraints are needed to improve the  significance of the goodness of fit. 
This means not only precise measurements on these observables, in particular at high$-z$, but also data in more redshift intervals. 
From our MCMC analysis, we observe that the data on $[f\sigma_8](z)$ was able to better constraint the parameters of the $f(R)$ model in both cases, thus results from upcoming surveys such as the DESI, SKA, and EUCLID telescopes  \cite{Maartens_2020} will be able to probe the growth of cosmic structures with precision to greatly improve the significance of this kind of analysis.

It is important to note that the conjoined analysis found in table  \ref{table_5}, even if with low statistical significance, was able to differentiate between different models inside the same $f(R)$ theory, and in particular with the same $n$ parameter, which we saw had high degeneracy in the $H(z)$ observables, and some degeneracy in $[f\sigma_8](z)$. Thus the proposal of \cite{Linder2017,Matsumoto2020} of breaking this degeneracy between models using the conjoined data is effectual.

\subsection{Remarks on stability and solar system test}

From the results of the MCMC runs, seen in \ref{table_4}, and from the constraints on the theory in \eqref{preconstraints}, that give $\lambda^{-1} < 0.21$ for the $n=1$ case, and  $\lambda^{-1} < 1.578$ for $n=2$, we see that no best fit for $\lambda$ in the $n=1$ case gives a stable model of the theory, while in the $n=2$ case the best fit with the $f\sigma_8$ dataset gives a fit that does not pass the de Sitter stability criteria. This once again shows some of the issues with the $n=1$ model.

As for the solar system test, the constraint comes mainly from the PPN parameter $\gamma$, which, from the Cassini probe, has to satisfy the constraint \cite{Negrelli_2020,Starobinsky2007}

\begin{equation}
    \label{cassini}
    |\gamma-1| < 4\times 10^{-4}.
\end{equation}

In a recent paper on solar system tests on $f(R)$ theories, \cite{Negrelli_2020} has shown that a model with the parameters $\lambda = \lambda^{-1} = 1$, $R_0 = 4.17H_0^2$ and $n=2$ easily passes the \eqref{cassini} constraint and the chameleon tests in the solar system. From the best fit parameters obtained in \ref{table_5}, we see that the best fit parameters for the $n=2$ are well inside the constraint obtained in \cite{Negrelli_2020}, since $R_0\lambda = 6\lambda(1-\Omega_{m0})H_0^2 < 4H_0^2$ for all the cases.

For the $n=1$ case, it is known since the original Starobinsky's work~\cite{Starobinsky2007} that for $n=1$ 
the model does not satisfy the constraints from the solar system test. 

\section{Conclusions and Final Remarks}\label{final-remarks}

In this paper, we have used the cosmic expansion and the structure growth data to constrain the free parameters of a relevant $f(R)$ gravity model, namely the Starobinsky model. 
We've used the bayesian based Monte Carlo Markov Chain method to explore the parameter space of this model from both sets of data individually and using a joint likelihood analysis. 
Furthermore, we also used the recent method proposed by Linder \cite{Linder2017} and Matsumoto et al. \cite{Matsumoto2020} to complement the result of  the MCMC statistical analysis. 
This novel approach, based on a joint data analyses from $H$ and $f\sigma_8$, has proven able to give further distinction between different models which  were shown to have degeneracy in the theoretical predictions of the $H(z)$ and $[f\sigma_8](z)$ observables in the range of the free parameters of the $f(R)$ model.

In the end, our results show that in the $f(R)$ Starobinsky model, for $n=1$ the model is not very well constrained but the data from $f\sigma_8$ gives a good fit to the joint data, where for $n=2$, the parameters are significantly better constrained and the joint data gives a great fit. The $\Lambda$CDM has the smallest $\chi^2$ value for the test, as expected. This joint analysis gives another observational result in motivating $n\geq 2$ in the $f(R)$ Starobinsky model.
However more --and more precise-- data are needed in order to determine the preferred model with good statistical significance. The joint binned data set still gives space to overfitting from the $\Lambda$CDM model and big error margins in the parameter space of the $f(R)$ models. Therefore, the Starobinsky model cannot be discarded as a possible alternative to the $\Lambda$CDM model, still passing all the observational tests for $n\geq 2$.

A possible continuation of this work would be to use this joint analysis to compare diverse  cosmologically motivated $f(R)$ models, such as the one in \cite{Hu2007} and the one in this paper  break the degeneracy between different $f(R)$ theories. The results of this paper show that this method is promising for further distinction between modified gravity models that are degenerate in certain cosmological probes. The extension of this analysis is left for future work.

\section*{Acknowledgements}
PB thanks CAPES for a PhD fellowship; AB thanks a CNPq fellowship. 
MC acknowledges the financial support from Programa de Capacitação Institucional (PCI/MCTI), 
that subsidized part of this work.\\
The authors are grateful for the computational facilities of the Observat\'orio Nacional.


\bibliographystyle{unsrt}  
\bibliography{main} 

\begin{thebibliography}{10}

\bibitem{Starobinsky2007}
A.~A. Starobinsky.
\newblock Disappearing cosmological constant in f(r) gravity.
\newblock {\em JETP Letters}, 86(3):157–163, Oct 2007.

\bibitem{Linder2017}
Eric~V. Linder.
\newblock {Cosmic growth and expansion conjoined}.
\newblock {\em Astroparticle Physics}, 86:41--45, 2017.

\bibitem{Planck:2015bue}
P.~A.~R. Ade et~al.
\newblock {Planck 2015 results. XIV. Dark energy and modified gravity}.
\newblock {\em Astron. Astrophys.}, 594:A14, 2016.

\bibitem{Planck2018}
N.~Aghanim et~al.
\newblock {Planck 2018 results. VI. Cosmological parameters}.
\newblock {\em Astron. Astrophys.}, 641:A6, 2020.
\newblock [Erratum: Astron.Astrophys. 652, C4 (2021)].

\bibitem{Kazantzidis:2018rnb}
Lavrentios Kazantzidis and Leandros Perivolaropoulos.
\newblock {Evolution of the $f\sigma_8$ tension with the Planck15/$\Lambda$CDM
  determination and implications for modified gravity theories}.
\newblock {\em Phys. Rev. D}, 97(10):103503, 2018.

\bibitem{Basilakos_2020}
Spyros Basilakos and Fotios~K. Anagnostopoulos.
\newblock Growth index of matter perturbations in the light of dark energy
  survey.
\newblock {\em The European Physical Journal C}, 80(3), Mar 2020.

\bibitem{Linder_2020}
Eric~V. Linder.
\newblock Limited modified gravity.
\newblock {\em Journal of Cosmology and Astroparticle Physics},
  2020(10):042–042, Oct 2020.

\bibitem{Velasquez_Toribio_2020}
A.~M. Velasquez-Toribio and Júlio~C. Fabris.
\newblock The growth factor parametrization versus numerical solutions in flat
  and non-flat dark energy models.
\newblock {\em The European Physical Journal C}, 80(12), Dec 2020.

\bibitem{Bonilla_2021}
Alexander Bonilla, Suresh Kumar, and Rafael~C. Nunes.
\newblock Measurements of $h_0$ and reconstruction of the dark energy
  properties from a model-independent joint analysis.
\newblock {\em The European Physical Journal C}, 81(2), Feb 2021.

\bibitem{Davari_2021}
Zahra Davari and Sohrab Rahvar.
\newblock Mog cosmology without dark matter and the cosmological constant.
\newblock {\em Monthly Notices of the Royal Astronomical Society},
  507(3):3387–3399, Aug 2021.

\bibitem{Linder}
Eric~V. Linder and Robert~N. Cahn.
\newblock Parameterized beyond-einstein growth.
\newblock {\em Astroparticle Physics}, 28(4):481 -- 488, 2007.

\bibitem{Benaoum}
H.~B. {Benaoum}, Weiqiang {Yang}, Supriya {Pan}, and Eleonora {Di Valentino}.
\newblock {Modified Emergent Dark Energy and its Astronomical Constraints}.
\newblock {\em arXiv e-prints}, page arXiv:2008.09098, August 2020.

\bibitem{Perenon15}
Louis Perenon, Federico Piazza, Christian Marinoni, and Lam Hui.
\newblock {Phenomenology of dark energy: general features of large-scale
  perturbations}.
\newblock {\em JCAP}, 11:029, 2015.

\bibitem{Perenon19}
Louis {Perenon}, Julien {Bel}, Roy {Maartens}, and Alvaro {de la Cruz-Dombriz}.
\newblock {Optimising growth of structure constraints on modified gravity}.
\newblock {\em JCAP}, 2019(6):020, June 2019.

\bibitem{Perenon20}
Louis Perenon, St\'ephane Il\'\i{}c, Roy Maartens, and Alvaro de~la
  Cruz-Dombriz.
\newblock {Improvements in cosmological constraints from breaking growth
  degeneracy}.
\newblock {\em Astron. Astrophys.}, 642:A116, 2020.

\bibitem{Batista14}
Ronaldo~C. Batista.
\newblock {Impact of dark energy perturbations on the growth index}.
\newblock {\em Phys. Rev. D}, 89(12):123508, 2014.

\bibitem{Nunes18}
Rafael~C. Nunes.
\newblock {Structure formation in $f(T)$ gravity and a solution for $H_0$
  tension}.
\newblock {\em JCAP}, 05:052, 2018.

\bibitem{Benisty18}
David Benisty and Eduardo~I. Guendelman.
\newblock {Unified dark energy and dark matter from dynamical spacetime}.
\newblock {\em Phys. Rev. D}, 98(2):023506, 2018.

\bibitem{Benisty19a}
Fotios~K. Anagnostopoulos, David Benisty, Spyros Basilakos, and Eduardo~I.
  Guendelman.
\newblock {Dark energy and dark matter unification from dynamical space time:
  observational constraints and cosmological implications}.
\newblock {\em JCAP}, 06:003, 2019.

\bibitem{Benisty19b}
David Benisty, Eduardo Guendelman, and Zbigniew Haba.
\newblock {Unification of dark energy and dark matter from diffusive
  cosmology}.
\newblock {\em Phys. Rev. D}, 99(12):123521, 2019.
\newblock [Erratum: Phys.Rev.D 101, 049901 (2020)].

\bibitem{Hu2007}
Wayne Hu and Ignacy Sawicki.
\newblock {Models of f(R) Cosmic Acceleration that Evade Solar-System Tests}.
\newblock {\em Physical Review D}, may 2007.

\bibitem{Clifton2011}
Timothy Clifton, Pedro~G. Ferreira, Antonio Padilla, and Constantinos Skordis.
\newblock {Modified Gravity and Cosmology}.
\newblock {\em Physics Reports}, jun 2011.

\bibitem{Martinelli2011}
Matteo Martinelli, Erminia Calabrese, Francesco De~Bernardis, Alessandro
  Melchiorri, Luca Pagano, and Roberto Scaramella.
\newblock Constraining modified gravitational theories by weak lensing with
  euclid.
\newblock {\em Physical Review D}, 83(2), Jan 2011.

\bibitem{Strauss}
Michael~A. Strauss and Jeffrey~A. Willick.
\newblock The density and peculiar velocity fields of nearby galaxies.
\newblock {\em Physics Reports}, 261(5):271 -- 431, 1995.

\bibitem{Basilakos12}
Spyros Basilakos.
\newblock The $\lambda$cdm growth rate of structure revisited.
\newblock {\em International Journal of Modern Physics D}, 21(07):1250064,
  2012.

\bibitem{Matsumoto2020}
Jiro Matsumoto, Teppei Okumura, and Misao Sasaki.
\newblock {New measures to test modified gravity cosmologies}.
\newblock {\em Journal of Cosmology and Astroparticle Physics}, 2020(7), jul
  2020.

\bibitem{DeFelice2010}
Antonio de~Felice and Shinji Tsujikawa.
\newblock {f (R) theories}.
\newblock {\em Living Reviews in Relativity}, 13, 2010.

\bibitem{Chen2019}
Yow-Chun Chen, Chao-Qiang Geng, Chung-Chi Lee, and Hongwei Yu.
\newblock {Matter Power Spectra in Viable $f(R)$ Gravity Models with Dynamical
  Background}.
\newblock {\em The European Physical Journal C}, 79(2):93, jan 2019.

\bibitem{Kopp_2013}
Michael Kopp, Stephen~A. Appleby, Ixandra Achitouv, and Jochen Weller.
\newblock Spherical collapse and halo mass function inf(r)theories.
\newblock {\em Physical Review D}, 88(8), oct 2013.

\bibitem{MOTOHASHI_2009}
Hayato Motohashi, Alexei~A. Starobinsky, and Junichi Yokoyama.
\newblock Analytic solution for matter density perturbations in a class of
  viable cosmological f(r) models.
\newblock {\em International Journal of Modern Physics D}, 18(11):1731--1740,
  nov 2009.

\bibitem{Cardone2012}
Vincenzo~F Cardone, Stefano Camera, and Antonaldo Diaferio.
\newblock An updated analysis of two classes of f(r) theories of gravity.
\newblock {\em Journal of Cosmology and Astroparticle Physics},
  2012(02):030–030, Feb 2012.

\bibitem{Hough2020}
Renier Hough, Amare Abebe, and Stefan Ferreira.
\newblock {Viability tests of f(R)-gravity models with Supernovae Type 1A
  data}, 2020.

\bibitem{Motohashi_2013}
Hayato Motohashi, Alexei~A. Starobinsky, and Jun'ichi Yokoyama.
\newblock Cosmology based on f(r) gravity admits 1 ev sterile neutrinos.
\newblock {\em Physical Review Letters}, 110(12), mar 2013.

\bibitem{Chudaykin_2015}
Anton~S. Chudaykin, Dmitry~S. Gorbunov, Alexei~A. Starobinsky, and Rodion~A.
  Burenin.
\newblock Cosmology based on f(r) gravity with (1) {eV} sterile neutrino.
\newblock {\em Journal of Cosmology and Astroparticle Physics},
  2015(05):004--004, 2015.

\bibitem{Sotiriou2010}
Thomas~P. Sotiriou and Valerio Faraoni.
\newblock f(r)theories of gravity.
\newblock {\em Reviews of Modern Physics}, 82(1):451–497, Mar 2010.

\bibitem{Motohashi_2011_1}
Hayato Motohashi, Alexei~A Starobinsky, and Junichi Yokoyama.
\newblock Future oscillations around phantom divide inf(r) gravity.
\newblock {\em Journal of Cosmology and Astroparticle Physics},
  2011(06):006--006, jun 2011.

\bibitem{Tsujikawa2007}
Shinji Tsujikawa.
\newblock Matter density perturbations and effective gravitational constant in
  modified gravity models of dark energy.
\newblock {\em Physical Review D}, 76(2), Jul 2007.

\bibitem{Jimenez_2002}
Raul Jimenez and Abraham Loeb.
\newblock Constraining cosmological parameters based on relative galaxy ages.
\newblock {\em The Astrophysical Journal}, 573(1):37--42, jul 2002.

\bibitem{Yang_2020}
Yingjie Yang and Yungui Gong.
\newblock The evidence of cosmic acceleration and observational constraints.
\newblock {\em Journal of Cosmology and Astroparticle Physics},
  2020(06):059--059, jun 2020.

\bibitem{1987MNRAS.227....1K}
Nick {Kaiser}.
\newblock {Clustering in real space and in redshift space}.
\newblock {\em MNRAS}, 227:1--21, July 1987.

\bibitem{Hamilton1992ApJ}
A.~J.~S. {Hamilton}.
\newblock {Measuring Omega and the Real Correlation Function from the Redshift
  Correlation Function}.
\newblock {\em APJL}, 385:L5, January 1992.

\bibitem{Avila19}
F.~Avila, C.~P. Novaes, A.~Bernui, E.~de~Carvalho, and J.~P.
  Nogueira-Cavalcante.
\newblock {The angular scale of homogeneity in the Local Universe with the SDSS
  blue galaxies}.
\newblock {\em Mon. Not. Roy. Astron. Soc.}, 488(1):1481--1487, 2019.

\bibitem{Marques20}
Gabriela~A. Marques and Armando Bernui.
\newblock {Tomographic analyses of the CMB lensing and galaxy clustering to
  probe the linear structure growth}.
\newblock {\em JCAP}, 05:052, 2020.

\bibitem{deCarvalho18}
E.~de~Carvalho, A.~Bernui, G.~C. Carvalho, C.~P. Novaes, and H.~S. Xavier.
\newblock {Angular Baryon Acoustic Oscillation measure at $z=2.225$ from the
  SDSS quasar survey}.
\newblock {\em JCAP}, 04:064, 2018.

\bibitem{deCarvalho20}
E.~de~Carvalho, A.~Bernui, H.~S. Xavier, and C.~P. Novaes.
\newblock {Baryon acoustic oscillations signature in the three-point angular
  correlation function from the SDSS-DR12 quasar survey}.
\newblock {\em Mon. Not. Roy. Astron. Soc.}, 492(3):4469--4476, 2020.

\bibitem{Nunes:2015xsa}
Rafael~C. Nunes, Ed\'esio~M. Barboza, Jr., Everton M.~C. Abreu, and
  Jorge~Ananias Neto.
\newblock {Probing the cosmological viability of non-gaussian statistics}.
\newblock {\em JCAP}, 08:051, 2016.

\bibitem{Alam21}
Shadab Alam, Marie Aubert, Santiago Avila, Christophe Balland, Julian~E.
  Bautista, Matthew~A. Bershady, Dmitry Bizyaev, Michael~R. Blanton, Adam~S.
  Bolton, Jo~Bovy, and et~al.
\newblock Completed sdss-iv extended baryon oscillation spectroscopic survey:
  Cosmological implications from two decades of spectroscopic surveys at the
  apache point observatory.
\newblock {\em Physical Review D}, 103(8), Apr 2021.

\bibitem{Avila21a}
F.~Avila, A.~Bernui, E.~de~Carvalho, and C.~P. Novaes.
\newblock {The growth rate of cosmic structures in the local Universe with the
  ALFALFA survey}.
\newblock {\em Mon. Not. Roy. Astron. Soc.}, 505(3):3404--3413, 2021.

\bibitem{Turnbull}
Stephen~J. Turnbull, Michael~J. Hudson, Hume~A. Feldman, Malcolm Hicken,
  Robert~P. Kirshner, and Richard Watkins.
\newblock Cosmic flows in the nearby universe from type ia supernovae.
\newblock {\em Monthly Notices of the Royal Astronomical Society},
  420(1):447–454, Dec 2011.

\bibitem{Huterer}
Dragan Huterer, Daniel~L. Shafer, Daniel~M. Scolnic, and Fabian Schmidt.
\newblock Testing $\lambda$ cdm at the lowest redshifts with sn ia and galaxy
  velocities.
\newblock {\em Journal of Cosmology and Astroparticle Physics},
  2017(05):015–015, May 2017.

\bibitem{Achitouv17}
I.~Achitouv, C.~Blake, P.~Carter, J.~Koda, and F.~Beutler.
\newblock Consistency of the growth rate in different environments with the
  6-degree field galaxy survey: Measurement of the void-galaxy and
  galaxy-galaxy correlation functions.
\newblock {\em Physical Review D}, 95(8), Apr 2017.

\bibitem{Beutler}
Florian Beutler, Chris Blake, Matthew Colless, D.~Heath Jones, Lister
  Staveley-Smith, Gregory~B. Poole, Lachlan Campbell, Quentin Parker, Will
  Saunders, and Fred Watson.
\newblock The 6df galaxy survey: $z\approx 0$ measurements of the growth rate
  and $\sigma_8$.
\newblock {\em Monthly Notices of the Royal Astronomical Society},
  423(4):3430–3444, Jun 2012.

\bibitem{Feix}
Martin Feix, Adi Nusser, and Enzo Branchini.
\newblock Growth rate of cosmological perturbations at $z\approx0.1$ from a new
  observational test.
\newblock {\em Physical Review Letters}, 115(1), Jun 2015.

\bibitem{Sanchez}
Ariel~G. Sánchez, Francesco Montesano, Eyal~A. Kazin, Eric Aubourg, Florian
  Beutler, Jon Brinkmann, Joel~R. Brownstein, Antonio~J. Cuesta, Kyle~S.
  Dawson, Daniel~J. Eisenstein, and et~al.
\newblock The clustering of galaxies in the sdss-iii baryon oscillation
  spectroscopic survey: cosmological implications of the full shape of the
  clustering wedges in the data release 10 and 11 galaxy samples.
\newblock {\em Monthly Notices of the Royal Astronomical Society},
  440(3):2692–2713, Apr 2014.

\bibitem{Blake12}
Chris Blake, Sarah Brough, Matthew Colless, Carlos Contreras, Warrick Couch,
  Scott Croom, Darren Croton, Tamara~M. Davis, Michael~J. Drinkwater, Karl
  Forster, and et~al.
\newblock The wigglez dark energy survey: joint measurements of the expansion
  and growth history atz< 1.
\newblock {\em Monthly Notices of the Royal Astronomical Society},
  425(1):405–414, Jul 2012.

\bibitem{Nadathur}
Seshadri Nadathur, Paul~M. Carter, Will~J. Percival, Hans~A. Winther, and
  Julian~E. Bautista.
\newblock Beyond bao: Improving cosmological constraints from boss data with
  measurement of the void-galaxy cross-correlation.
\newblock {\em Physical Review D}, 100(2), Jul 2019.

\bibitem{Chuang16}
Chia-Hsun Chuang, Francisco Prada, Marcos Pellejero-Ibanez, Florian Beutler,
  Antonio~J. Cuesta, Daniel~J. Eisenstein, Stephanie Escoffier, Shirley Ho,
  Francisco-Shu Kitaura, Jean-Paul Kneib, and et~al.
\newblock The clustering of galaxies in the sdss-iii baryon oscillation
  spectroscopic survey: single-probe measurements from cmass anisotropic galaxy
  clustering.
\newblock {\em Monthly Notices of the Royal Astronomical Society},
  461(4):3781–3793, Jun 2016.

\bibitem{Pezzotta17}
A.~Pezzotta, S.~de~la Torre, J.~Bel, B.~R. Granett, L.~Guzzo, J.~A. Peacock,
  B.~Garilli, M.~Scodeggio, M.~Bolzonella, U.~Abbas, and et~al.
\newblock The vimos public extragalactic redshift survey (vipers).
\newblock {\em Astronomy \& Astrophysics}, 604:A33, Jul 2017.

\bibitem{Aubert20}
Marie Aubert, Marie-Claude Cousinou, Stéphanie Escoffier, Adam~J. Hawken,
  Seshadri Nadathur, Shadab Alam, Julian Bautista, Etienne Burtin, Arnaud
  de~Mattia, Héctor Gil-Marín, Jiamin Hou, Eric Jullo, Richard Neveux,
  Graziano Rossi, Alex Smith, Amélie Tamone, and Mariana~Vargas Magaña.
\newblock The completed sdss-iv extended baryon oscillation spectroscopic
  survey: Growth rate of structure measurement from cosmic voids, 2020.

\bibitem{Wilson}
Michael~J. Wilson.
\newblock Geometric and growth rate tests of general relativity with recovered
  linear cosmological perturbations, 2016.

\bibitem{Zhao}
Gong-Bo Zhao, Yuting Wang, Shun Saito, Héctor Gil-Marín, Will~J Percival,
  Dandan Wang, Chia-Hsun Chuang, Rossana Ruggeri, Eva-Maria Mueller, Fangzhou
  Zhu, and et~al.
\newblock The clustering of the sdss-iv extended baryon oscillation
  spectroscopic survey dr14 quasar sample: a tomographic measurement of cosmic
  structure growth and expansion rate based on optimal redshift weights.
\newblock {\em Monthly Notices of the Royal Astronomical Society},
  482(3):3497–3513, Oct 2018.

\bibitem{Okumura}
Teppei Okumura, Chiaki Hikage, Tomonori Totani, Motonari Tonegawa, Hiroyuki
  Okada, Karl Glazebrook, Chris Blake, Pedro~G. Ferreira, Surhud More, Atsushi
  Taruya, and et~al.
\newblock The subaru fmos galaxy redshift survey (fastsound). iv. new
  constraint on gravity theory from redshift space distortions at $z\approx
  1.4$.
\newblock {\em Publications of the Astronomical Society of Japan}, 68(3):38,
  Apr 2016.

\bibitem{Moresco2017}
M.~Moresco and F.~Marulli.
\newblock {Cosmological constraints from a joint analysis of cosmic growth and
  expansion}.
\newblock {\em Monthly Notices of the Royal Astronomical Society: Letters},
  471(1):L82--L86, 2017.

\bibitem{Riess:2019cxk}
Adam~G. Riess, Stefano Casertano, Wenlong Yuan, Lucas~M. Macri, and Dan
  Scolnic.
\newblock {Large Magellanic Cloud Cepheid Standards Provide a 1\% Foundation
  for the Determination of the Hubble Constant and Stronger Evidence for
  Physics beyond $\Lambda$CDM}.
\newblock {\em Astrophys. J.}, 876(1):85, 2019.

\bibitem{NB20}
Rafael~C. Nunes and Armando Bernui.
\newblock {BAO signatures in the 2-point angular correlations and the Hubble
  tension}.
\newblock {\em Eur. Phys. J. C}, 80(11):1025, 2020.

\bibitem{Anagnostopoulos}
Fotios~K. Anagnostopoulos and Spyros Basilakos.
\newblock Constraining the dark energy models with h(z) data: An approach
  independent of h0.
\newblock {\em Physical Review D}, 97(6), Mar 2018.

\bibitem{Foreman_Mackey_2013}
Daniel Foreman-Mackey, David~W. Hogg, Dustin Lang, and Jonathan Goodman.
\newblock emcee: The mcmc hammer.
\newblock {\em Publications of the Astronomical Society of the Pacific},
  125(925):306–312, Mar 2013.

\bibitem{Marques19}
Gabriela~A. Marques, Jia Liu, Jos\'e Manuel~Zorrilla Matilla, Zolt\'an Haiman,
  Armando Bernui, and Camila~P. Novaes.
\newblock {Constraining neutrino mass with weak lensing Minkowski Functionals}.
\newblock {\em JCAP}, 06:019, 2019.

\bibitem{Planck2016}
P.~A.~R. Ade, N.~Aghanim, M.~Arnaud, M.~Ashdown, J.~Aumont, C.~Baccigalupi,
  A.~J. Banday, R.~B. Barreiro, J.~G. Bartlett, and et~al.
\newblock Planck2015 results.
\newblock {\em Astronomy \& Astrophysics}, 594:A13, Sep 2016.

\bibitem{Avila21b}
F.~Avila, A.~Bernui, R.~C. Nunes, E.~de~Carvalho, and C.~P. Novaes.
\newblock {The homogeneity scale and the growth rate of cosmic structures}.
\newblock {\em Monthly Notices of the Royal Astronomical Society}, 2021.

\bibitem{Koivisto2006}
Tomi Koivisto.
\newblock {Matter power spectrum in f(R) gravity}.
\newblock {\em Physical Review D - Particles, Fields, Gravitation and
  Cosmology}, 73(8):1--5, 2006.

\bibitem{Maartens_2020}
Jan-Albert Viljoen, José Fonseca, and Roy Maartens.
\newblock Constraining the growth rate by combining multiple future surveys.
\newblock {\em Journal of Cosmology and Astroparticle Physics},
  2020(09):054–054, Sep 2020.

\bibitem{Negrelli_2020}
Carolina Negrelli, Lucila Kraiselburd, Susana~J. Landau, and Marcelo Salgado.
\newblock Solar system tests and chameleon effect in f(r) gravity.
\newblock {\em Physical Review D}, 101(6), mar 2020.

\end{thebibliography}
\end{document}